\documentclass[twocolumn,preprintnumbers,amsmath,amssymb]{revtex4}

\usepackage{graphicx}
\usepackage{dcolumn}
\usepackage{bm}

\newcommand{\be}{\begin{equation}}
\newcommand{\ee}{\end{equation}}
\newcommand{\bfig}{\begin{figure}}
\newcommand{\efig}{\end{figure}}

\begin{document}      

\title{Spin entropy as the likely source of enhanced thermopower in $\rm\bf Na_xCo_2O_4$\footnote{Nature {\bf 423}, 425 (2003).}} 
\author{Yayu Wang$^1$, Nyrissa S. Rogado$^2$, R. J. Cava$^{2,3}$, and N. P. Ong$^{1,3}$}      
\affiliation{
$^1$Department of Physics, 
$^2$Department of Chemistry, 
$^3$Princeton Materials Institute,\\
Princeton University, Princeton, New Jersey 08544. 
}

\date{January 20th, 2003}      


\maketitle                   

{\bf In an electric field, the flow of electrons in a solid produces an entropy current in addition to the familiar charge current.  This Peltier effect underlies all 
thermoelectric refrigerators.  The upsurge in thermoelectric cooling applications has led to a search for more efficient Peltier materials and to renewed 
theoretical interest in how electron-electron interaction may enhance the thermopower $Q$ of materials such as the transition-metal 
oxides~\cite{Mahan,Beni,Kotliar,Chaikin}.  An important factor in this enhancement is the electronic spin entropy, which is 
predicted~\cite{Chaikin,Kwak,KwakChaikin} to dominate the entropy current.  However, the crucial evidence for the spin-entropy term, namely its 
complete suppression in a longitudinal magnetic field, has not been reported to date.  Here we report evidence for such suppression in the layered oxide 
$\rm Na_xCo_2O_4$, and present evidence that it is a strong-correlation effect.   The field suppression of $Q$ provides a rare, unambiguous example of 
how strong interaction effects can qualitatively alter electronic behavior in a solid.  We discuss implications for the search for 
better Peltier materials.}

In a thermopower experiment, a temperature gradient $-\nabla T$ drives the diffusion of charge carriers to the cooler end of the sample.  The charge 
accumulation leads to a net electric field $\bf E$ which determines the thermopower (or Seebeck coefficient) $Q = E/|\nabla T|$~\cite{Ziman}.  In 
conventional metals, the flow of electrons induced by $-\nabla T$ is nearly cancelled by a parallel flow of holes, so that $Q$ is strongly reduced by the 
factor $T/T_F$ from the `natural value' $k_B/e\sim 86\; \mu$V/K, where $k_B$ is Boltzmann's constant, $e$ the electron charge and $T_F$ the Fermi 
temperature ~\cite{Ziman}.   Moreover, as the currents are indifferent to the spins, a longitudinal magnetic field ${\bf H} ||-\nabla T$ has virtually no effect on 
$Q$.  In materials with strong electron-electron interaction, however, the spin degrees are predicted to produce a large contribution of the Heikes 
form~\cite{Beni,Chaikin,KwakChaikin} 
\be
Q\rightarrow  \frac{\mu}{eT} = -\frac{\sigma}{e},			
\label{Heikes}
\ee
where $\mu$ is the chemical potential, and $\sigma$ (the entropy per electron) equals $k_B\ln(g_sg_c)$ with $g_s$ and $g_c$ the spin and configuration 
degeneracies, respectively.  From Eq. \ref{Heikes}, the enhancement to $Q$ from spin entropy is of the order $k_B/e$, but it is suppressed to zero if the 
spin degeneracy is lifted in a magnetic field ($g_s\rightarrow 1$).  

Terasaki, Sasago, and Uchinokura~\cite{Terasaki} recently reported the discovery that $Q$ in $\rm Na_xCo_2O_4$ (at 300 K) is $\sim$10 times larger 
than in typical metals, and suggested that spin-fluctuations, as in heavy-fermion systems, may play a key role.  Despite Hall effect~\cite{Terasaki}, heat 
capacity~\cite{Ando} and nuclear spin-resonance~\cite{NMR} experiments, however, the enhancement in $Q$ remains puzzling.  Experiments have not 
ruled out either strong-interaction theories involving spin configurations~\cite{Maekawa}, or conventional interpretations based on small Fermi 
Surfaces~\cite{Singh}.  By investigating the thermopower and magnetization in both longitudinal and transverse $H$, we show that $\rm Na_xCo_2O_4$ 
is in fact a strong-correlation system in which the spin entropy term accounts for virtually all of $Q$ at 2 K and a dominant 
fraction at 300 K.

\begin{figure}[h]			
\includegraphics[width=8cm]{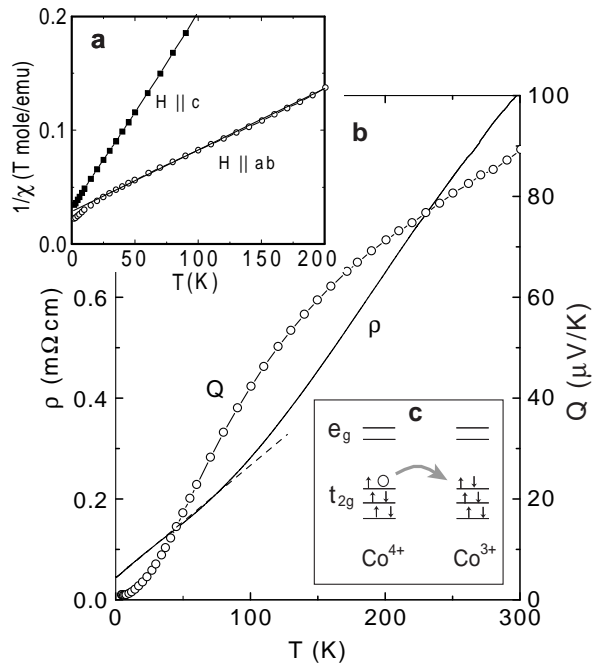}
\caption{\label{rho} The temperature (T) dependence of magnetic and transport properties of single-crystal $\rm Na_xCo_2O_4$ and electronic states in 
the Co ions.  Panel (a) shows the $T$ dependence of the inverse susceptibilities $\chi_{ab}$ and $\chi_c$ measured in a 5-Tesla field applied in-plane 
and along the $c$-axis, respectively.  In Panel (b), the in-plane thermopower $Q$ (open symbols) and resistivity $\rho$ (solid curve) are plotted vs. $T$.  
The $T$-linear dependence of  $\rho$ below 100 K is indicated by the dashed line.  The inset (c) is a sketch of the low-spin $t_{2g}$ states in the 
Co$^{4+}$ and Co$^{3+}$  ions.  The hopping of a hole (arrow) is accompanied by the transfer of a spin-$\frac12$ excitation. 
}
\end{figure}

In $\rm Na_xCo_2O_4$, the Co ions occupy the sites of a two-dimensional (2D) triangular lattice.  Chemical arguments and band-structure 
calculations~\cite{Singh} show that the oxygen $2p$ states lie far below the Co-$3d$ states and the chemical potential falls within the band formed from 
$t_{2g}$ states in Co.  Hence the electrons donated by Na ions are distributed among the Co ions, a fraction $\delta$ of which are in the Co$^{4+}$ state 
while the rest ($1-\delta$) are Co$^{3+}$.  From $x$ = 1.36 (determined by ICP chemical analysis of our crystals), we find $\delta =$ 0.32.  The variation of 
the susceptibilities $\chi_{ab}$ and $\chi_c$ vs. temperature $T$ (Fig. \ref{rho}a) fit accurately to the Curie-Weiss form $\sim (T+\theta)^{-1}$, implying the 
existence of large local moments that are antiferromagnetically coupled.  The plot of $1/\chi_{ab}$ gives $\theta = 55 \pm 5$ K and a moment $\sim 
0.87\;\mu_B$ averaged over all Co ions ($\mu_B$ is the Bohr magneton).  This experimentally constrains the Co$^{4+}$ ions to be in the low-spin state 
with spin $S = \frac12$ and moment $\sqrt{3}\mu_B$, while the Co$^{3+}$ ions are in their low-spin $S = 0$ state (Fig. 1c), in agreement with Ref. 
~\cite{NMR}.

\begin{figure}				
\includegraphics[width=8cm]{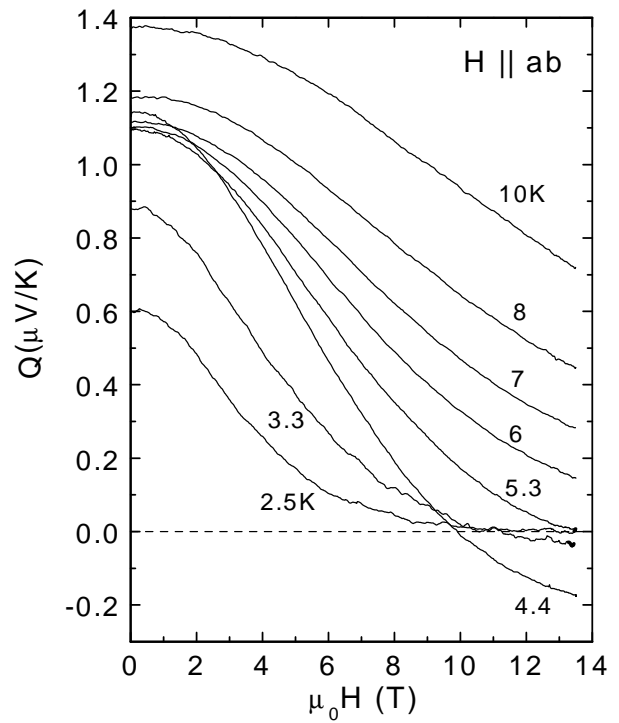}
\caption{\label{Qplane}   The in-plane thermopower $Q$ vs. an in-plane $\bf H\parallel(-\nabla T)$ at selected $T$.  At 4.4 K, $Q$ asymptotes at high fields 
to a negative floor value ($-0.25 \mu$V/K).  At the lowest $T$ (2.5 K), $Q$ approaches a value unresolved from zero when $H$ exceeds 8 T.  The applied 
temperature difference $\delta T$ is less than 0.3 K below 4 K, and ~0.5 K above 4 K.  By using phosphor bronze as the voltage leads, we have 
eliminated background contributions to $Q$ (Supplementary Information).
}
\end{figure}

The in-plane resistivity $\rho$  displays a $T$-linear variation below 100 K that is suggestive of strong-correlation behavior (dashed line in Fig. \ref{rho}b).  
As noted in Ref.~\cite{Terasaki}, $Q$ (open circles) increases with $T$ to values much larger than in conventional metals.  However, it is the effect of a 
magnetic field on $Q$ that is the most telling.  The pronounced effect of an in-plane field $H$ on $Q$ is shown in Fig. \ref{Qplane} (${\bf H} || -\nabla T$).   
Stringent precautions against spurious contributions to $Q$ were adopted (see Supplementary Information).  Even at moderately high $T$ (30 K), $Q$ 
decreases significantly in a 14-Tesla field.  Lowering the temperature makes the field-suppression of $Q$ more pronounced, until at 4 K, $Q$ changes 
sign and saturates near 12 T to a 'floor' value equal to -0.25 $\mu$V/K.  With further decrease in $T$, the floor value itself monotonically decreases to zero.  
The trace at 2.5 K shows that $Q$ decreases monotonically to zero in a field of 8 T, and remains at zero over an extended field 
range.

\begin{figure}				
\includegraphics[width=8cm]{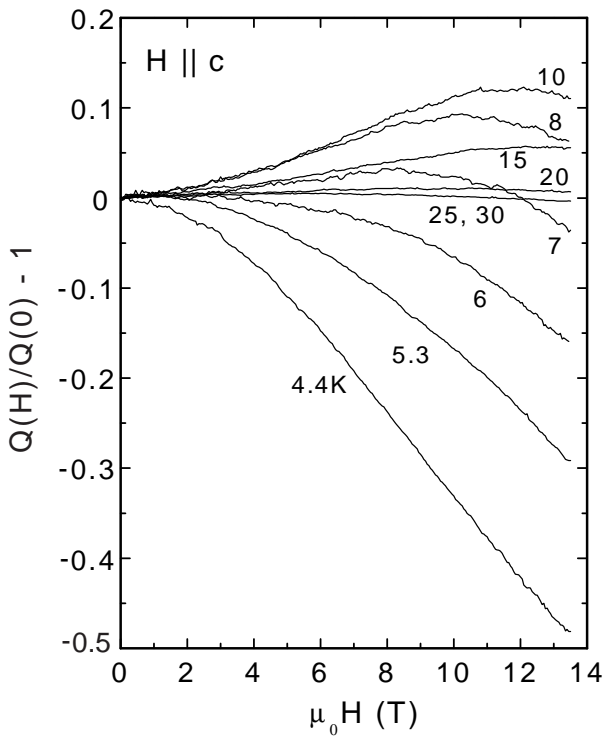}
\caption{\label{Qtrans}  The relative change in $Q$ versus a transverse field $\bf H || c$ in $\rm Na_xCo_2O_4$.  At 6 K and lower, $Q$ decreases 
monotonically with $H$, but at higher temperatures, $Q$ initially increases in weak field before decreasing at larger $H$.    
}
\end{figure}

The strong effect of field on $Q$ is also observed with $\bf H || c$ (normal to plane), except that, above 6 K, we observe an additional term that produces 
an increase in $Q$ in weak fields.   At 4.2 K, $Q$ displays the same field profile as above, but with a proportionately weaker decrease (Fig. \ref{Qtrans}).  
As the sample is warmed to 10 K and higher, the relative change in $Q$ is initially positive in low fields.  At higher fields, the negative trend becomes 
again apparent.  The overall pattern in transverse field is consistent with a positive contribution superposed on the term intrinsic to the spin degrees 
uncovered in the longitudinal field.

The evidence that the $H$ dependence of $Q$ arises from the spin degrees of freedom derives from both the transport and susceptibility experiments.  
As an in-plane field couples only to the spin degrees, but not the orbital degrees, the strong $H$ dependence of $Q$ in Fig. \ref{Qtrans} must arise from 
the effect of field on the spins.  In the geometry with $\bf H || c$, we reason that the non-monotonic field dependence of $Q$ also arises entirely from the 
effect of $H$ on the spin degrees, as the carrier mean-free-path $\ell\sim 100\;$\AA ~estimated from $\rho$ is far too short to produce any orbital effect in 
a field of 14 Tesla.  Comparing curves at 4 K in Figs. 2 and 3, we infer that a similar decrease in $Q$ requires a $c$-axis field that is 1.4 times larger than 
the in-plane field.  The close agreement between this ratio and the ratio of the bulk susceptibilities $\chi_{ab}/\chi_c$ at 4 K strongly suggests that the field 
anisotropies in the magnetization and thermopower experiments have the same origin.  The field suppression of $Q$ is larger with the field in-plane 
compared with the field along the c-axis because it is easier to align the spins with an in-plane field.    
In conventional metals, a longitudinal field merely changes the relative populations of the spin-up (+) and spin-down (-) electrons without significantly 
altering the entropy current in each spin population~\cite{Ziman}.  The complete suppression displayed in Fig. \ref{Qplane} is incompatible with a 
conventional band picture.

\begin{figure}				
\includegraphics[width=8cm]{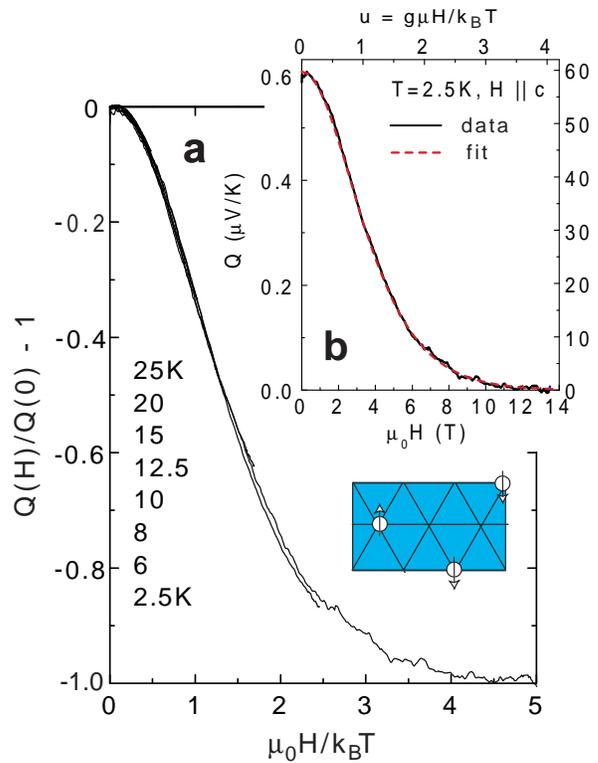}
\caption{\label{fit}  The suppression of the thermopower by an in-plane magnetic field in $\rm Na_xCo_2O_4$.  In Panel (a) the normalized $Q$ is plotted 
versus   $\mu_0 H/k_BT$.  Curves measured at various $T$ collapse to a universal curve.  The inset in (a) shows the -$\frac12$ excitations in an inert 
background of Co$^{3+}$ ions.  Panel (b) compares the curve at 2.5 K (solid curve) with the expression in Eq. \ref{SvsH}  (dashed 
curve).
}
\end{figure}

We next discuss how the field suppression may be understood within a strong-correlation scenario.  As shown in Fig. \ref{fit}a (insert), in the 2D triangular 
lattice of the Co ions, a fraction  $\delta = \frac13$ are in the Co$^{4+}$ state in which $S = \frac12$, while the rest are in the Co$^{3+}$ state with $S=0$ 
(the higher-lying $e_g$ orbitals are unoccupied).  The elementary charge transport process is the hopping of a hole from Co$^{4+}$ to Co$^{3+}$ (Fig. 
\ref{rho}c).  A large on-site repulsion excludes double-occupancy of a site by the holes.  Because this process converts the Co$^{4+}$ ($S = \frac12$) to a 
Co$^{3+}$ ($S = 0$) and vice versa, we also transfer a spin-$\frac12$ along with the hole, which implies a transfer of spin entropy $\sigma = k_B\ln 2$ in 
zero field.   The excitations are holes carrying charge +1$e$ and spin-$\frac12$ moving in a sea of Co$^{3+}$ sites, which are magnetically and 
electrically inert.  An E field leads to the heat current $J_Q = nvk_BT \ln 2$ as well as the charge current $J = nev$ ($n$ is the hole concentration and $v$ 
the average velocity).  As the ratio $J_Q/J$ (called the Peltier coefficient $\Pi$) equals the product $QT$, we obtain the spin-entropy part of $Q$ given in 
Eq. \ref{Heikes}.  Moreover, because the spin-entropy current is in the direction of the charge current, this contribution to $Q$ is positive (hole-like) as 
observed.  As $T$ decreases below  $\sim$50 K (the antiferromagnetic exchange energy  $J_{AF}\sim\theta$), the spin entropy must decrease rapidly 
with incipient magnetic ordering to vanish at $T$ = 0.  When $T\ll \theta$, the majority of the spins become antiferromagnetically correlated with their 
neighbors, leaving only a small fraction $T/\theta$  to behave as free spins.  In an E-field, this residual population of free spins carries the spin entropy 
although all the holes remain itinerant.  This implies that the ratio $J_Q/J$ decreases steeply below $\theta$, consistent with the observed behavior of $Q$ 
in Fig. \ref{rho}.

In a magnetic field, the residual spin entropy $\sigma(H)$ is further decreased, eventually reaching 0 when $H$ is strong enough to completely lift the 
2-fold degeneracy.  Modeling the residual free spins as non-interacting spins with the Land\'{e} g-factor $g$, we calculate the field dependence of the 
entropy as
\begin{equation}
\frac{\sigma (H,T)}{\sigma(0)} =  \frac{\ln(2\cosh u) - u\tanh u}{\ln 2}, \quad (u = \frac{g\mu_B H}{2k_BT}).
\label{SvsH}
\end{equation}
This dependence should apply to the component of $Q$ derived from the spin entropy.  

As shown in Fig. \ref{fit}a, the curves of $Q$ in Fig. \ref{Qplane} between 2.5 and 25 K can be collapsed onto a universal curve when plotted against the 
ratio $H/T$.  Using the curve at 2.5 K (which extends to the largest $u$), we find that Eq. \ref{SvsH} provides a remarkably close fit if the $g$-factor is 
$2.2\pm 0.1$  (Fig. \ref{fit}b).  We view the close fit (achieved with a reasonable $g$) as confirming evidence that the entropy current observed in the 
thermopower indeed derives entirely from the spin-$\frac12$ excitations, and the complete suppression of $Q$ reflects the removal of the spin 
degeneracy by the applied field.

We remark that, at each $T$, the observed $Q$ is comprised of several distinct contributions.  Application of a strong longitudinal $H$ completely 
suppresses only the spin-entropy term, leaving the remaining terms unaffected (the configuration entropy $k_B\ln g_c$, for instance) which are revealed as 
the floor value at high fields.  In Fig. \ref{Qplane}, these terms amount to $-0.25 \mu$V/K at 4.4 K and rapidly vanish at lower $T$.  We emphasize that the 
spin entropy contribution to $Q$ is not restricted to low $T$.  Measurements at high fields (30 T) show that $Q$ is suppressed by ~20 $\%$ at 30 K.  
Assuming that, when $T\gg\theta$, the spin contribution asymptotes to $\frac{k_B}{e}\ln 2 \sim 60\;\mu$V/K, we find that it constitutes $\sim\frac23$ of the 
total thermopower at 300 K, accounting for most the enhanced thermopower.

The spin-entropy enhancement may be widespread in the transition-metal oxides.  A key ingredient inferred from our measurements is the motion of 
charge tied to local moments of spin $\frac12$ in a background of Co$^{3+}$ ions that are magnetically inert (diamagnetic).  Among the large class of 
cobalt oxides, the Co ions are nearly always in their low-spin state.  In the Co oxides that can be made conducting, therefore, the thermopower is likely to 
be dominated by spin-entropy terms, suggesting that this is a promising class of materials to search for improved Peltier materials.

We acknowledge support from the U.S. National Science Foundation (NSF).  Some of the measurements were performed at the U.S. National High 
Magnetic Field Laboratory, Tallahassee, Fl., which is supported by NSF and the state of Florida. We thank Scott Hannahs for 
technical assistance.

\newpage
~~
\newpage
\centerline{\bf Supplementary Material}
In high-field measurements of thermopower $Q$, there is an important source of background contribution, namely the thermopower $Q_w$ of the voltage 
lead at the warm end of the sample.  It is essential to determine separately the field dependence (if any) of $Q_w$.  After testing several candidate metals, 
we have found that phosphor bronze is nearly ideal because its $Q_w$ is very small and shows no measurable field dependence at 5 K 
up to 30 Tesla.

\begin{figure}				
\includegraphics[width=7cm]{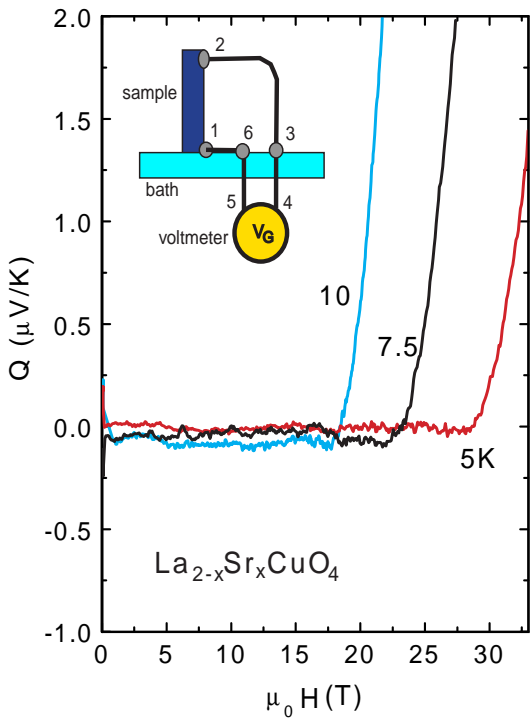}
\caption{\label{LSCO}  (Supplementary information)  The thermoelectric signal versus $H$ with single-crystal $\rm La_{2-x}Sr_xCuO_4$ ($x$ = 0.17) 
serving as the reference sample.  At each $T$, the voltmeter recording $V_G$ is $H$-independent and close to zero until $H$ exceeds the vortex solid 
melting field $H_m(T)$ ($\sim$30 T at 5 K).  At $H_m(T)$, $V_G$ increases abruptly reflecting the large thermopower in the vortex liquid state.  The flat 
behavior of $V_G$ for $H<H_m(T)$ verifies that the thermopower of phosphor bronze $Q_w$ is field independent at least up to $H_m(T)$.  The insert 
shows the closed loop 1-2-3-4-5-6 used in the derivation of Eq. \ref{S1}.
}
\end{figure}

In a thermopower experiment, one end of the sample (2) is held at temperature $T+\delta T$ while the other end 1 is anchored at the bath temperature 
$T$.  The $E$-field generated by the Seebeck effect is determined from the potential difference of the voltage leads electrically connected to the two 
sample ends.  However, the Seebeck coefficient $Q_w$ of the voltage lead at the warmer end invariably contributes to the voltage reading.  To see this, 
we consider potential changes around the close loop 1-2-3-4-5-6 (insert in Fig. \ref{LSCO}).  The voltage leads 23 and 16 are made of phosphor bronze 
while the long twisted pair 34 and 56 are made of copper.  With $V_G = V_4 - V_5$, we have
\begin{eqnarray}
-\int_1^2 \;d{\bf l\cdot E} &-& \int_2^3 \;d{\bf l\cdot E}_w - \int_3^4\;d{\bf l\cdot E}_{w'} \nonumber\\
- \int_5^6 \;d{\bf l\cdot E}_{w'} & -& \int_6^1 \;d{\bf l\cdot E}_w - V_G = 0,		
\label{S1}
\end{eqnarray}
where ${\bf E}_w$ and ${\bf E}_{w'}$ are the $E$-field in phosphor bronze and in Cu, respectively.   The 1st and 2nd terms are the potential drops across 
the sample, and the `warm' voltage lead 23, respectively.  The 3rd and 4th terms are the potential drops along the long, twisted pair of copper wires 
extending from the sample chamber to the preamplifier at room temperature.  The 5th term, the potential drop in the `cool' voltage lead 16, vanishes as 
nodes 1 and 6 are anchored at the bath temperature.  In a carefully designed apparatus, the 3rd and 4th terms cancel to zero.  However, the 2nd term can 
never be eliminated because the same temperature drop $\delta T$ falls across the sample and the voltage wire 23, i.e. $T_3 = T_1$. Using ${\bf E} = 
Q\nabla T$, and  $T = T_2-T_1$ in Eq. \ref{S1}, we have
\be
V_G = (Q-Q_w)  \delta T.  				
\label{S2}
\ee

In high-resolution experiments, the field dependence of $Q_w$ in the voltage lead material (phosphor bronze) must be measured separately.  Ideally, if 
$Q$ is rigorously zero up to the maximum $H$ of interest, $V_G$ would just equal -$Q_w T$.  In an extreme type II superconductor, the thermopower is 
strictly zero in a magnetic field if the vortices remain pinned (in the solid state).  We used an optimally-doped crystal of $\rm La_{2-x}Sr_xCuO_4$ ($x$ = 
0.17) in which the melting field of the vortex solid $H_m(T)$ has been studied extensively.  For fields $H<H_m(T)$, the voltmeter reading is then $V_G  = 
-Q_w \delta T$ at each $T$.  Hence any field variation in $Q_w$ can be observed directly.  Figure \ref{LSCO} shows that as $H$ is increased, $V_G$ 
increases abruptly at $H_m(T)$ reflecting the increase in $Q$.  However, below $H_m(T)$, the voltmeter reading is virtually independent of $H$, 
confirming that $Q_w$ in phosphor bronze is field independent (at least, up to $H_m(T)$) and very small at these temperatures.  With this test, we verify 
explicitly that the field dependences shown in Figs. 2 and 3 entirely derive from $\rm Na_xCo_2O_4$.

\end{document}